\documentclass[traditabstract]{aa}
\usepackage[utf8]{inputenc}
\usepackage[activate={true,nocompatibility},final,tracking=true,kerning=true,spacing=true]{microtype}

\usepackage{paralist}
\usepackage{amsmath}
\usepackage{natbib}
\usepackage{graphicx}
\usepackage{epsfig}
\usepackage{txfonts}
\usepackage{color}
\usepackage{multirow}
\usepackage{amssymb}
\usepackage{epsfig}
\usepackage{xspace}
\bibpunct{(}{)}{;}{a}{}{,} 
\usepackage[plainpages=false]{hyperref}

\newcommand{\vela}{Vela~X$-$1\xspace}
\newcommand{\gx}{GX~301$-$2\xspace}
\newcommand{\src}{4U~1907$+$09\xspace}
\newcommand{\rt}{Rayleigh–Taylor\xspace}
\newcommand{\kh}{KHI\xspace}

\title{Supergiant, fast, but not so transient 4U 1907+09}
\author{V.\,Doroshenko\inst{1}, A.\,Santangelo\inst{1}, L.\,Ducci\inst{1}, D.\,Klochkov\inst{1}}	
\institute{Institut für Astronomie und Astrophysik, Sand 1, 72076 Tübingen, Germany}

\begin{document}

\bibliographystyle{aa}
\abstract{We have investigated the dipping activity observed in the high-mass
X-ray binary \object{4U 1907+09} and shown that the source continues to pulsate in the ``off''
state, noting that the transition between the ``on'' and ``off'' states may be
either dip-like or flare-like. This behavior may be explained in the framework
of the ``gated accretion'' scenario proposed to explain the flares in supergiant
fast X-ray transients (SFXTs). We conclude that \src might prove to be a missing
link between the SFXTs and ordinary accreting pulsars.}

\authorrunning{V. Doroshenko et al.}
\maketitle

\section{Introduction}The persistently active high-mass X-ray binary system
\src, discovered in the third \emph{Uhuru} survey \citep{giacconi71}, consists
of an X-ray pulsar with a spin period of $\sim 437.5$ \citep{makishima84} and a
highly reddened companion in an eccentric $e\sim0.28$ orbit with a period of
$P\sim8.3753$\,d \citep{zand98}. Optical and infrared observations
\citep{cox05,nespoli08} suggest an O8-O9 Ia type supergiant donor with
$L\sim5\times10^5L_{\odot}$ and a mass-loss rate of about
$7\times10^{-6}M_{\odot}\rm{yr}^{-1}$. The supergiant nature of the companion
and the strong X-ray variability indicate that accretion most likely proceeds
from a stellar wind. The distance to the source is estimated to be between 2 and
6\,kpc \citep{cox05,nespoli08}, therefore, the X-ray luminosity of the source is
uncertain. Assuming a compromise distance of 4\,kpc, the observed persistent
flux of $\sim10^{-10}$\,erg\,cm$^{-2}$\,s$^{-1}$ \citep{zand98} implies a
luminosity of about $10^{36}\,{\rm erg\,s}^{-1}$.

The broadband X-ray spectrum of \src, similar to other accreting pulsars, can be
modeled with an absorbed cut-off power law. Based on \emph{Ginga} data,
\cite{mihara95} reported a cyclotron resonance scattering feature (CRSF) at
19\,keV, which was later confirmed with \emph{BeppoSAX} \citep{cusumano98}, who
also reported a harmonic at $\sim39$\,keV and a narrow iron emission line at
$6.4$\,keV. The CRSF harmonic, however, was not detected either in \emph{RXTE}
\citep{zand98} and \emph{Suzaku} \citep{Rivers} observations.

One signature feature of \src is its dipping activity. During the dips, the
X-ray flux drops abruptly by a decade or more for a period of few minutes to
several hours \citep{zand98,asca,Rivers}. Similar behavior has been reported for
other HMXBs, notably Vela~X$-$1 \citep{kreyken_vela,me_vela} and GX~301$-$2
\citep{kreyken10}. However, \src spends a substantial fraction of time (up to
60\%, \citealp{sahiner}) in the ``off'' state, whereas dips are rare events in
other systems. Unlike \object{Vela X-1} and \object{GX 301-2}, no pulsations were detected during the
short dips of \src observed with \emph{RXTE} \citep{zand98}. \cite{asca} were able to
detect pulsed emission during an extended low-flux episode observed with
\emph{ASCA}. Two scenarios have been invoked to explain the dips: 1) obscuration
of the neutron star by a dense wind clump, and 2) cessation of the accretion
owing a wind density drop. Several authors have suggested that in most cases
there is no evidence of an increase in the absorption column during the dips
\citep{zand98,asca,Rivers}, so the former scenario is unlikely.

In this study, we have examined the nature of the dips in the light-curve of
\src in great detail, using recent \emph{Suzaku} and \emph{RXTE} observations.
First, we compared the flaring behavior of the source with the one of other
HMXBs, and then we attempted to link the observed phenomenology with the flaring
activity typical of the SFXTs. We affirm that similarities of \src between both
normal wind accreting pulsars and SFXTs suggest the source might constitute a
missing link between the two classes.

\section{Observations and facts}
\begin{figure*}[t]
	\centering
		\includegraphics[width=1.0\textwidth]{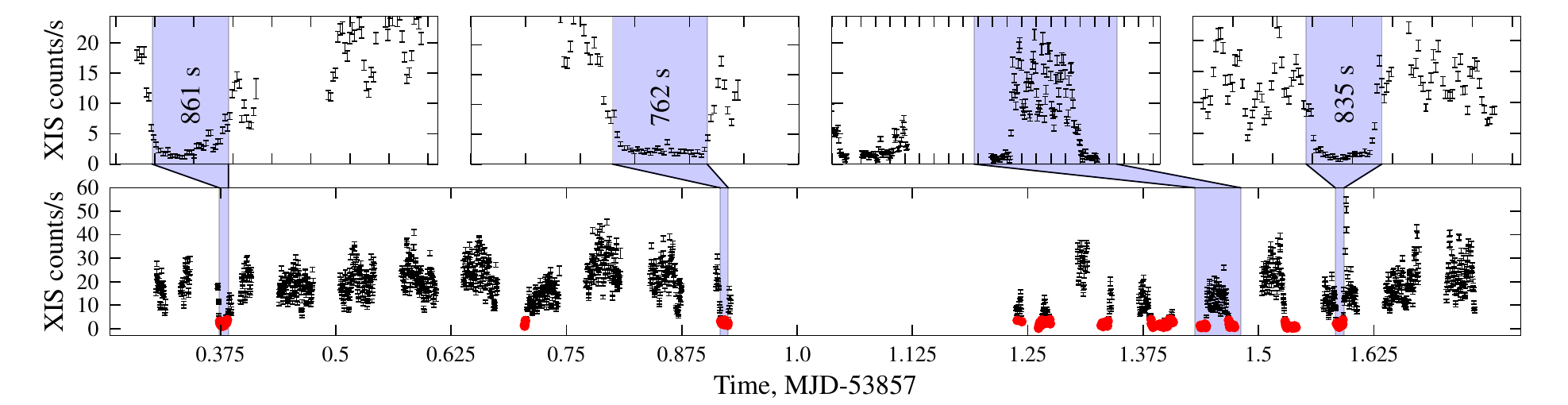}
	\caption{The observation-long XIS lightcurve (background-subtracted with 4 XIS units combined) and close-up of some of the ``off'' states, all of which are marked with red points in the lightcurve.}
	\label{fig:obslc}
\end{figure*}

\paragraph{Pulsations during the ``off''-states.} For a deeper understanding of
the nature of the source, it is essential to clarify whether the residual X-ray
emission of the ``off''-states is pulsed or not. This is indeed the case for
Vela~X$-$1 \citep{me_vela} and GX~301$-$2 \cite{kreyken10}. As already
mentioned, \hbox{\cite{zand98}} did not detect pulsations during the dips of
\src using \emph{RXTE} observations, however, the source lies close to the
galactic ridge, and therefore systematic effects related to a stronger
background might affect observations with collimating instruments. On the other
hand, \emph{ASCA} observations hint at the pulsations during the two 10\,ks-long
extended low-flux episodes \citep{asca} when the source flux was comparable to
the ``off''-states. It is unclear, however, whether the pulsations were
significant and whether such a low-flux episode could be caused by the same
mechanism as the short dips. To clarify this, we analyzed a 123\,ks
\emph{Suzaku} observation (ID~401057010 from May 2006) of the source. We
repeated the analysis carried out by \cite{Rivers} who used this observation to
study the behavior of the source in the normal flux state, focusing, however,
exclusively on the source properties during the ``off-states''.

\begin{figure}[t]
	\centering
		\includegraphics[width=0.5\textwidth]{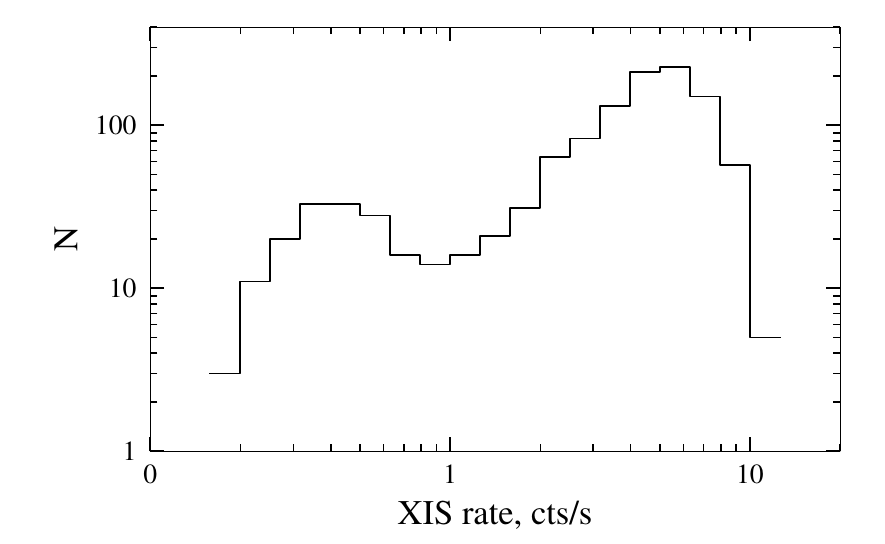}
	\caption{Count-rate distribution in the background-subtracted XIS lightcurve (44\,s bin size). Two roughly log-normally distributed peaks corresponding to ``off'' and ``on'' states can be identified.}
	\label{fig:hist}
\end{figure}

\emph{Suzaku} is equipped with two instruments: the X-ray spectrometer (XIS),
consisting of focusing four units and a collimating hard X-ray detector (HXD).
To increase the signal-to-noise ratio, we only used XIS data, combining all four
units for timing analysis. For data reduction, we used the HEASOFT 6.11 analysis
package, and the set of calibration files v.20111109. The XIS lightcurve of the
entire observation is presented in Fig.~\ref{fig:obslc}. Several dip episodes
are observed throughout the entire observation. Similar to \vela
\citep{me_vela}, the observed count rate roughly follows a log-normal
distribution for both the ``on'' and ``off'' states, with different mean values
as is also evident from the lightcurve (see Fig.~\ref{fig:hist}). Here we define
``off''-states as the abrupt flux drops to a level of $\sim1$\,cts/s per XIS
unit for at least one spin cycle of the pulsar. Note, that it is expected to
find that the ``off''-states contribute to the \emph{tail} of the ``on'' flux
distribution rather than forming a separate \emph{peak}, should they come from
random fluctuations in wind density or velocity.

\begin{figure}[t]
	\centering
		\includegraphics[width=0.5\textwidth]{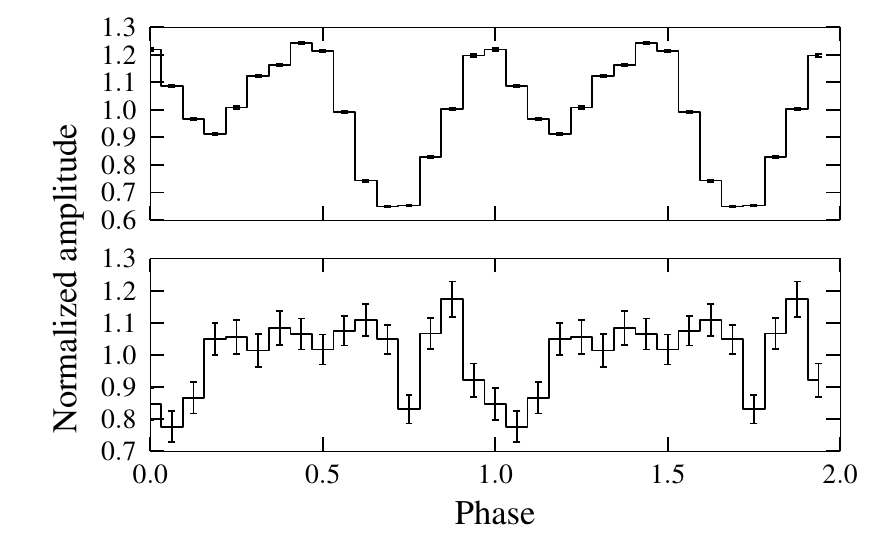}
	\caption{Background subtracted ``on'' (top) and ``off'' (bottom) XIS pulse profiles in 1-10\,keV energy range folded with the same parameters.}
	\label{fig:pp}
\end{figure}
\begin{figure}[t]
	\centering
		\includegraphics[width=0.5\textwidth]{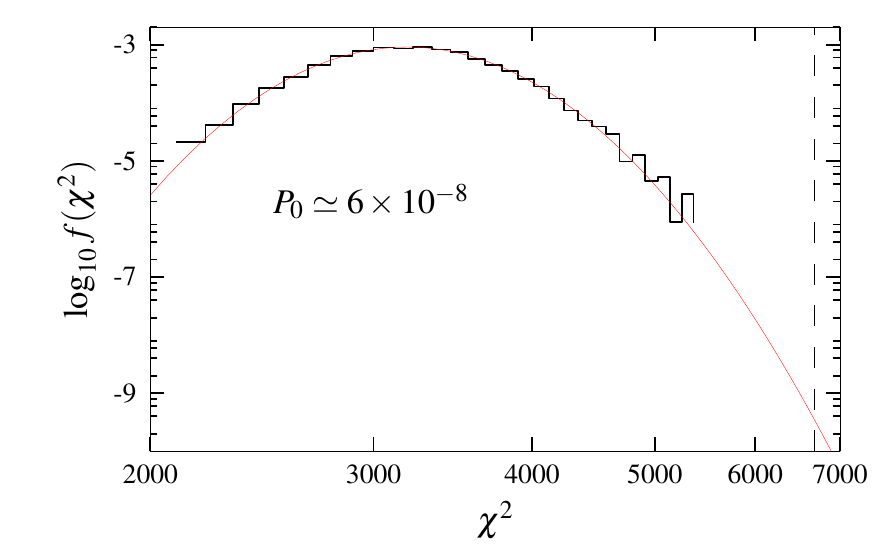}
	\caption{The distribution of squared deviations in the synthetic pulse profiles
from the constant (black) and the best-fit probability density function to the
distribution (red). Dashed line indicates the observed squared deviation for the
``off'' pulse profile.}
	\label{fig:pdf}
\end{figure}

For the timing analysis, the lightcurve was corrected for the orbital motion in
the solar and in the binary system using the orbital parameters by
\cite{zand98}. Phase-coherent timing analysis, using the XIS and HXD
light-curves, yielded a constant pulse period of P=441.10(5)\,s, consistent with
the value reported by \cite{Rivers}.

A blind search for significant pulsations in the ``off'' lightcurve is
challenging because it only contains a few pulse cycles separated by many gaps
of various lengths, which alias with the intrinsic variability and result in
very noisy periodograms if conventional algorithms are used. On the other hand,
folding the ``off'' lightcurve with the period found from ``on'' lightcurve
produces a pulse profile with a similar shape (see Fig.~\ref{fig:pp}), which is
also similar to the \emph{ASCA} folded lightcurve for the extended low-flux
episode (see Fig.~5 of \citealt{asca}). To justify the significance of the
pulsations one can estimate the probability of getting the observed deviation
from a constant over the pulse profile by chance (for known pulse-period and
pulse shape). To do this we simulated $10^{4}$ modified ``off'' lightcurves with
flux randomly distributed with the same average value and dispersion as the
observed ``off'' flux. We then folded the simulated lightcurves with observed
period and calculated the deviation of the resulting pulse profile from a
constant. These turn out to be distributed log-normally, so one can estimate the
probability of getting the deviation equal to or larger than observed. We find
that it is fairly low at $P_0\sim6\times10^{-8}$ as summarized in
Fig.~\ref{fig:pdf}, i.e. the observed pulsation is significant.

\vspace{-0.3cm}
\paragraph{``Interdips'' or flares} It is interesting to observe that the
fraction of time the source spends in the ``off'' state is so long that
occasionally two consecutive dips are separated by just a short ``on''
interval of a comparable length to that of the dips. This interval would be
considered a transient flaring episode, should only the part of the lightcurve
shown in the third upper panel of Fig.~\ref{fig:obslc} be observed. We
searched for similar events in the source light-curves in \emph{RXTE} archive,
that had extensively monitored the source. Searching for uninterrupted
pointings, which fully covered intervals between consecutive dips, we found
the two cases presented in Fig.~\ref{fig:interdips}. We would like to
emphasize that the sharp transition between the ``off'' and ``on'' states
produces distinct flare-like events in the lightcurve.

Out of the \emph{RXTE} archival data \citep{sahiner} another relevant
observational fact emerges: the probability of observing the source in the ``off''-state
depends on the orbital phase and is significantly lower at periastron. 

\section{Discussion}
\begin{figure}[t] \centering
\includegraphics[width=0.5\textwidth]{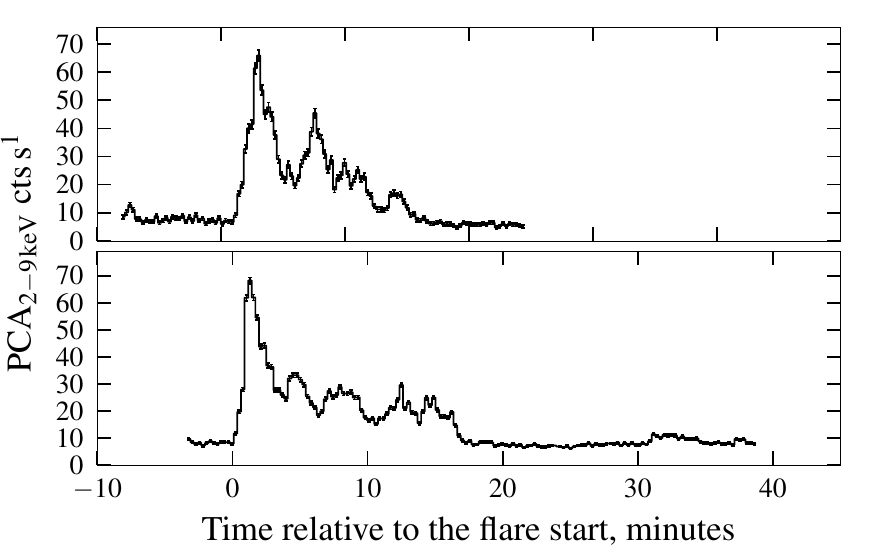}
\caption{Two SFXT-like
flares observed by \emph{RXTE} and centered respectively at $\sim$MJD\,55169.5 and
$\sim$MJD\,51074.49. In the latter case the source remains ``off'' after a gap
in the data for $\sim3$\,h.} \label{fig:interdips}
\end{figure}

The discussion section begins with a short summary of the observed phenomenology:
\begin{itemize}
	\item \src exhibits a dipping activity associated with sudden drops in flux by an order of magnitude or more on a timescale that is comparable to the spin period. Dips last several minutes to several hours before the flux recovers to the ``on'' value on a similar timescale. 
	\item \src is not unique and a similar behavior has been observed in other HMXBs, notably \vela and \gx.
	\item In the ``off'' state, accretion does not cease completely, forming a distinct peak in the flux distribution while the source continues to pulsate.
		\item The pulse profile changes significantly during the ``off''-state.
	\item The probability of observing the source in the ``off'' state depends on the orbital phase and is lower at periastron.
	\item The fraction of time \src spends in the ``off''-state is long enough to make inter-dip intervals occasionally appear as flares.
\end{itemize}
First, emphasis should be placed on the phenomenological similarity between \src
and the SFXTs. The inter-dips or flares shown in Fig~\ref{fig:interdips}
exhibit timescales and amplitudes similar to the smaller flares observed from the
SFXTs. Indeed, we have measured an average
unabsorbed flux of $\sim4.5\times10^{-11}{\rm erg\, cm}^{-2}{\rm s}^{-1}$ and
$\sim4.5\times10^{-10}{\rm erg\, cm}^{-2}{\rm s}^{-1}$ for the ``off'' and
``on'' fluxes respectively,  implying
$L_x\simeq8.6\times10^{34-35}{\rm erg\,s}^{-1}$ for an assumed distance of
4\,kpc, so is comparable to values observed during the flaring periods in SFXTs
\citep{sguera,Bozzo:2008p2039}.

On the other hand, \src in ``off'' state is brighter than the SFXTs in deep
quiescence, and it has shorter ``off'' duty cycle. The latter depends, however,
on the orbital phase, hence on the distance from the companion. One could
anticipate, therefore, that the source would remain ``off'' longer should the
system be wider. On the other hand, the duty cycle of SFXTs is also known to
depend on the orbital phase, and some of them leave the deep quiescence and only
start to flare in the vicinity of the periastron \citep{sidoli11}. Making such a
system more compact would extend the flaring period and prevent if from going to
deep quiescence, thus making it similar to the \src with its relatively bright
``off'' states and lower dynamical range than the prototypical SFXTs.
Luminosity-related changes of the pulse profiles similar to ones in \src have
also been observed in SFXT, at least in one case \citep{romano09}. \src seems to
exhibit, therefore, properties that are intermediate between the SFXTs and the
persistent systems.

The dipping activity in \src is not unique among wind-accreting pulsars. As
discussed by \cite{kreyken_vela} for \vela, the drop in X-ray flux must be
intrinsic rather than due to eclipses by a wind clump obscuring the neutron
star. Since no ``off'' flux was known to emerge from \vela and \src,
\cite{kreyken_vela} and \cite{sahiner} have suggested that the centrifugal
inhibition might be responsible for the cessation of emission. However, residual
pulsed flux has been detected in \vela \citep{me_vela}, suggesting that a change
in the accretion regime might be responsible for the flux drop. Following
\cite{burnard83} and \cite{Bozzo:2008p2039}, \cite{me_vela} proposed for \vela that, while
normally the accretion proceeds via \rt instability (RTI), this switches off
during the dips, and Kelvin–Helmholtz instabilities (\kh) become relevant. Note
that the switch between two accretion regimes naturally explains the observed
bimodal flux distribution. The same scenario may also apply to \src.

It is interesting to discuss why just a few X-ray pulsars do switch ``off''. For
SFXTs \cite{Bozzo:2008p2039} argue that a strong magnetic field is required to
explain the observed luminosity swings and the characteristic accretion rates in
different accretion regimes. They estimate that RTIs become inefficient for
mass-loss rates of the optical companion below:
\begin{equation}
	\dot{M}_{-6}\leq280P^{-3}_{s3}a^2_{10d}\upsilon_8R^{5/2}_{M10}\left[1+16R_{a10}/(5R_{M10})\right]^{-3/2}
\end{equation}
Here, $R_{\rm G10}$ and $R_{\rm M10}$ are the capture and magnetosphere radius,
respectively, in units of $10^{10}$\,cm; $a_{10d}=P_{10d}^{2/3}M_{30}^{1/3}$ is
function of the orbital period and of the total mass of the system; $\upsilon_8$
is the relative velocity between the wind and the neutron star in units of
$10^8$\,cm\,s$^{-1}$; and $P_{s3}$ is the spin period in units of 1000\,s.
Assuming for \src the parameters reported by \cite{cox05}, we obtain an upper
limit for the magnetosphere size of $R_{M10}\leq0.6$. Routinely observed regime
transitions imply, however, that the $R_{M10}$ is close to this value, i.e.
large as well.

\cite{Bozzo:2008p2039} also provide an estimate for the leak rate and the
corresponding X-ray luminosity if plasma enters the magnetosphere via \kh. For
\src the estimate is
\begin{eqnarray*}
& & L_{\rm KH} \simeq G M_{\rm NS} \dot{M}_{\rm KH}/R_{\rm NS}\simeq\nonumber \\ 
& & 10^{35}\eta_{\rm KH} R_{\rm M10}^3
(1+16 R_{\rm G10}/(5 R_{\rm M10}))^{3/2} 
\frac{\sqrt{\rho_{\rm i}/\rho_{\rm e}}}{
1+\rho_{\rm i}/\rho_{\rm e}} ~{\rm erg ~s}^{-1}\nonumber \\
\label{eq:lkh} 
\end{eqnarray*}
Where $\rho_{\rm i,e}$ are the densities within and outside of the
magnetosphere. According to \cite{Bozzo:2008p2039}, $\eta_{\rm KH}\sim 0.1$,
and the density ratio is estimated to be between
\begin{equation*}
	\frac{\sqrt{\rho_{\rm i}/\rho_{\rm e}}}{
	1+\rho_{\rm i}/\rho_{\rm e}}=\begin{cases}
	0.3\eta_{\rm KH} h^{-1} R_{\rm M10}^{3/2} P_{\rm s3}^{-1}\\
	0.1 \eta_{\rm KH} h^{-1} R_{\rm M10}^{1/2} v_{8}
\end{cases}
\end{equation*}
where $h$ is the fractional height of the area where the plasma and the magnetic
field coexist, in units of the total thickness of the \kh unstable layer
\citep{burnard83}. In the case of \src, for the observed ``off-state''
luminosity of $\sim8\times10^{34}\,{\rm erg\,s}^{-1}$, $R_{M10}\geq0.25$ can be
estimated if the \kh unstable layer is relatively thin ($h\sim0.05$), or
$R_{M10}\geq 0.55$, if $h\sim1$, as suggested by \cite{burnard83}. The latter is
in good agreement with the value deduced from stability criteria. For the
observed X-ray luminosity, this implies a magnetized neutron star with
$B\sim10^{13}$\,G.

\cite{kreyken_vela} proposed that the magnetosphere might grow enough to inhibit
the accretion even for moderately magnetized neutron star if the wind density in
the vicinity of the neutron star drops by factor of $\sim1000$. This would allow
reconciliation of the ``off''-states with a magnetic field intensity of
$B\simeq2\times10^{12}$\,G like the one deduced from the observed CRSF energy.
The winds of young supergiants are known to be structured, and hydrodynamical
simulations have predicted a wind density contrast as high as $10^{4}$
\citep{oskinova}. These predictions are not confirmed by observations, however,
because the observed X-ray flux typically varies at most by a factor of 10.
Moreover, the ``off-state'' duty cycle of \src depends on the orbital phase
\citep{sahiner}; i.e., the fluctuations in wind density which trigger the
``off''-states must be comparable to the ones associated with the orbital
motion. The average wind density in \src is expected in the CAK model
\citep{castor} to vary just by a factor of $\sim5$ along the orbit, which is
consistent with the observed orbital lightcurve. The ``off''-states in \src are
therefore most likely triggered by minor fluctuations in wind density, so a
large magnetosphere would imply a magnetic field for the neutron star of
$B\sim10^{13}$\,G. Such a strong magnetic field might help also in explaining
the long spin-period of \src, although its complicated evolution definitively
deserves a dedicated study.

The observed CRSF energy in \src implies, however, an order of magnitude weaker
field $B\sim1.5\times10^{12}$\,G in the line-forming region. Considering that
the details of the magnetosphere-plasma interaction are far from being fully
understood, and that there is significant uncertainty in the distance to the
source (which reflects in the estimate of the accretion rate and consequently of
the magnetosphere radius), it may well be that the scenario outlined above will
also hold for a weaker field. However, \src is a very unusual HMXB, and we tend
to believe that there is something distinctive behind its dipping behavior, and
the strong magnetic field of the neutron star is a natural candidate. As
previously discussed for the case of \vela and \gx, the CRSF energy provides an
estimate for the magnetic field in a line-forming region, and if the line forms
far above surface of the neutron star, it might underestimate the surface field
by orders of magnitude. The dipole component of the field changes with $\sim
R^{-3}$, so scattering in the upper parts of the accretion column, or in the
accretion stream at heights comparable to the neutron star radius atop the polar
caps may help explain the observed CRSF energies even for magnetar-like surface
fields. Note, that the complicated observed shapes of the pulse profiles in
bright accreting pulsars cannot be explained with the direct emission from
compact polar caps \citep{Kraus:2003}, and this possibility should be considered
irrespectively of the surface field.

In \src, the luminosity is probably too low to form an extended accretion column
\citep{Lyubarsky1988}, however, scattering of X-ray photons from the polar caps
in the upper accretion stream may be important. This was discussed by
\cite{Kraus:2003}, who find that scattering in the upper accretion flow might
affect the formation of the pulse-profile and spectra in a major way and, in
fact, be responsible for the majority of the hard photons coming from the
source. The magnetic field in scattering region might already be compatible with
the one estimated from the observed CRSF energy. \cite{Nishimura:2008p2969}
consider the CRSF formation in an accretion column with a height of several
kilometers and suggest that line-like features are still present in the
spectrum. As we discussed for \gx \citep{Doroshenko:2010p3661}, the orientation
of the emission region with respect to the observer will also affect line
formation because only part of the emission or scattering region will be
visible.

Another scenario that could explain the difference between SFXTs and normal
HMXBs, invokes properties of the stellar wind. Recently, \cite{ikshanov} have
suggested that the picture of interaction of the accretion flow and the
magnetosphere outlined by \cite{Bozzo:2008p2039} should be altered if the wind
plasma itself is magnetized, and this might help explain the long pulse periods
in some pulsars. We note here that the wind magnetic field might also affect the
stability of the accretion, although how still needs to be clarified. In this
scenario the strong magnetic field of the \emph{optical} companion could, in
principle, be responsible for the difference between the ordinary HMXBs, the
intermediate systems like \src, and the SFXTs. Such a scenario requires further
theoretical investigation, while systematic survey of the magnetic fields of the
primaries of different source classes may be carried out already now to verify
whether they differ significantly.

\begin{acknowledgements}
VD and AS thank the Deutsches Zentrums für Luft- und Raumfahrt (DLR) and
Deutsche Forschungsgemeinschaft (DFG) for financial support (grant
DLR~50~OR~0702).
\end{acknowledgements}

\vspace{-0.8cm}\bibliography{auto_clean}	\vspace{-0.3cm}\end{document}